\documentclass[aps,pra,twocolumn,floatfix]{revtex4}
\usepackage{amsfonts, amssymb, amsmath,graphicx}
\usepackage{subfigure}

\begin{document}

\title{Ground-state phases of ultracold bosons with Rashba-Dresselhaus spin-orbit coupling}

\author{Tomoki Ozawa}
\author{Gordon Baym}
\affiliation{
Department of Physics, University of Illinois at Urbana-Champaign, 1110 W. Green St., Urbana, Illinois 61801, USA
}%

\date{\today}

\def\del{\partial}
\def\p{\prime}
\def\simge{\mathrel{%
         \rlap{\raise 0.511ex \hbox{$>$}}{\lower 0.511ex \hbox{$\sim$}}}}
\def\simle{\mathrel{
         \rlap{\raise 0.511ex \hbox{$<$}}{\lower 0.511ex \hbox{$\sim$}}}}
\newcommand{\feynslash}[1]{{#1\kern-.5em /}}

\begin{abstract}
We study ultracold bosons in three dimensions with an anisotropic Rashba-Dresselhaus spin-orbit coupling.
We first carry out the exact summation of ladder diagrams for the two-boson $t$ matrix at zero energy.
Then, with the $t$ matrix as the effective interaction,
we find the ground-state phase diagrams of bosons in mean field, as a function of the spin-orbit coupling, the anisotropy, and the scattering lengths between particles in the same and in different pseudospin states.
The resulting phase diagrams have much richer structures than those obtained using mean-field couplings, exhibiting three different phases: a plane wave condensate, a striped condensate, and an unstable phase.
The differences between the present approach using the $t$ matrix compared to using mean-field couplings is significant for large scattering lengths, large spin-orbit-coupling strength, or small anisotropy.

\end{abstract}

\maketitle

\section{Introduction}
Spin-orbit coupling plays a crucial role in a variety of physical systems ranging from atoms and nuclei  to topological insulators \cite{Hasan2010} and spintronics \cite{Zutic2004}.
Recently, the prospect of realizing spin-orbit coupling in ultracold atomic systems has led to increased interest not only in fermions but also bosons for this purpose \cite{Dalibard2010}.
Currently proposed and realized schemes to produce spin-orbit coupling in ultracold atomic systems
create effective non-Abelian gauge fields for atoms \cite{Dalibard2010,Ruseckas2005,Zhu2006,Liu2009,Juzeliunas2010,Campbell2011,Lin2011} which take the form of
Rashba-Dresselhaus spin-orbit couplings, familiar in semiconductor physics \cite{Rashba1960,Dresselhaus1955}.
A system of spin-orbit coupled ultracold bosons was recently realized by Lin et al. \cite{Lin2011}.

The problem of ultracold bosons with Rashba-Dresselhaus spin-orbit coupling has been considered within
mean-field, where the interparticle interactions are assumed 
to be
independent of momenta \cite{Stanescu2008,Wu2008,Wang2010,Jian2011,Yip2011,Zhang2011}.  These studies predicted that the ground states can be either a ``plane-wave" Bose-Einstein condensate  (BEC) of particles in a single momentum state, or a ``striped" BEC involving a coherent superposition of two different momenta.
However, recent studies indicate that,  beyond mean field, the effective interaction has a
qualitatively different structure, resulting, for example, in the absence of interaction between particles scattering in the zero total-momentum channel \cite{Gopalakrishnan2011,Ozawa2011} and the prediction that a BEC involving only a single momentum of particles is not favorable in an isotropic Rashba field \cite{Gopalakrishnan2011}.

In this study we 
delineate the ground state phase diagram of bosons taking the effective interaction described in terms of the full $t$ matrix rather than mean-field coupling.
We consider bosons with an isotropic or anisotropic Rashba-Dresselhaus spin-orbit coupling, $\sim \kappa (\sigma_x p_x + \eta \sigma_y p_y)$, where $\kappa$ is the spin-orbit coupling strength, $\mathbf{p}$ is the particle momentum,
and $\eta$ determines the $x$-$y$ anisotropy.  In our previous paper \cite{Ozawa2011}, we showed that the effective interaction can be renormalized in terms of physical scattering lengths and does not depend on the ultraviolet cutoffs.  Here, we extend this calculation, and carry out the exact summation of ladder diagrams for the $t$ matrix of bosons scattering at zero energy in three dimensions.
Then we reduce the Hamiltonian, with the $t$ matrix as the effective interaction, to a Nozi\`eres model \cite{Nozieres1995}, from which we determine the
ground-state phases as functions of the anisotropies in the $\kappa a_{aa}$-$\kappa a_{ab}$ plane,
where $a_{aa}$ and $a_{ab}$ are the scattering lengths between the same and different pseudospin states, respectively.
We show that there are generally three phases: a plane wave BEC, a striped BEC with two different momenta, and an unstable phase where the effective interaction is attractive
\footnote{That there are only three possible phases is, as pointed out in Ref.~\cite{Ho1996}, a consequence of the system having just two pseudospin components, which when dressed by a spin-orbit interaction can be miscible (the striped phase), immiscible (the plane wave phase), or unstable. See Ref.~\cite{Ho2010} for a discussion of the experiment of Ref.~\cite{Lin2011}.}.
The phase diagrams (Figs.\ref{zeroto075} and \ref{zero999to1} below) are substantially different and richer in structure than those predicted using mean-field couplings with the bare couplings
replaced by $4\pi \hbar^2 a_{ij}/m$,
especially when $\kappa a_{aa}$ and $\kappa a_{ab}$ are large, or $\eta$ is close to unity.
In the isotropic limit $\eta = 1$,  the plane wave BEC does not appear (Fig.~\ref{zero999to1}).  We describe how the phase diagram evolves continuously with anisotropy, from $\eta = 0$ where spin-orbit coupling is present only in the $x$ direction, to an isotropic spin-orbit coupling $\eta = 1$.
We find that, in the vicinity of $\eta = 1$, the phase diagrams are logarithmically sensitive to small changes in $\eta$, a structure we analyze by expanding in the anisotropy about $\eta = 1$.
We also find that a BEC with negative scattering lengths can  be stabilized in the presence of spin-orbit coupling, when the scattering lengths are sufficiently large in magnitude (Fig.~\ref{negativeeta05}).

Our analysis can be directly compared with proposed experimental schemes \cite{Ruseckas2005,Zhu2006,Liu2009,Juzeliunas2010,Campbell2011} when the scattering lengths are all equal; a good approximation for the three $F=1$ hyperfine states of $^{87}$Rb.  We discuss in this situation how the phases of bosons in the ground state evolve with varying anisotropy and scattering length (Fig.~\ref{su2_phases}).  In the following we take $\hbar=1$.


\section{Hamiltonian}

We consider bosons in three dimensions with a Rashba-Dresselhaus spin-orbit coupling, described by the Hamiltonian 
\begin{align}
	\mathcal{H}
	&=
	\sum_{\mathbf{p}}
	\begin{pmatrix}
	a_\mathbf{p}^\dagger & b_\mathbf{p}^\dagger
	\end{pmatrix}
	\left[
	\frac{p^2 + \kappa^2}{2m}I + \frac{\kappa}{m} (\sigma_x p_x + \eta \sigma_y p_y)
	\right]
	\begin{pmatrix}
	a_\mathbf{p} \\
	b_\mathbf{p}
	\end{pmatrix}
	\notag \\
	&+
	\frac{1}{2V}
	\sum_{\mathbf{p}_1 + \mathbf{p}_2 = \mathbf{p}_3 + \mathbf{p}_4}
	\left(
	g_{aa} a_{\mathbf{p}_4}^\dagger a_{\mathbf{p}_3}^\dagger a_{\mathbf{p}_2} a_{\mathbf{p}_1}
	\right.
	\notag \\
	&\hspace{1cm}\left.
	+
	g_{bb} b_{\mathbf{p}_4}^\dagger b_{\mathbf{p}_3}^\dagger b_{\mathbf{p}_2} b_{\mathbf{p}_1}
	+
	2g_{ab}\ a_{\mathbf{p}_4}^\dagger b_{\mathbf{p}_3}^\dagger b_{\mathbf{p}_2} a_{\mathbf{p}_1}
	\right).
\end{align}
As in our previous paper \cite{Ozawa2011}, $m$ is the atomic mass, $V$ is the volume of the system, $a_\mathbf{p}$ annihilates an atom the pseudospin state $a$ with momentum $\mathbf{p}$, and $b_\mathbf{p}$ annihilates an atom in the pseudospin state $b$ with momentum $\mathbf{p}$; the $\sigma_x$ and $\sigma_y$ are the usual Pauli matrices between the internal states, and 
$I$ is the two-by-two identity matrix.  We take the coupling $\kappa$ to be positive.  The $g_{aa}$, $g_{bb}$, and $g_{ab}$ are the bare $s$-wave couplings between $a$-$a$, $b$-$b$, and $a$-$b$ particles.
When the system is isotropic, $\eta = 1$, the Hamiltonian completely reduces to that considered previously \cite{Ozawa2011}.  In practice the effective Hamiltonian, in the basis in which the coupling has the Rashba-Dresselhaus form, can contain terms that do not conserve the individual number of particles of each species ($a$-like and $b$-like); we ignore such terms at this point, and return to this issue below. 

To diagonalize the single-particle part of the Hamiltonian we introduce operators $\alpha_\mathbf{p}$ and $\beta_\mathbf{p}$ defined by
\begin{align}
	\begin{pmatrix}
	\alpha_\mathbf{p} \\
	\beta_\mathbf{p}
	\end{pmatrix}
	=
	\frac{1}{\sqrt{2}}
	\begin{pmatrix}
	1 & -e^{-i\phi} \\
	1 & e^{-i\phi}
	\end{pmatrix}
	\begin{pmatrix}
	a_\mathbf{p} \\
	b_\mathbf{p}
	\end{pmatrix},
\end{align}
where $\phi$ is the angle of $(p_x, \eta p_y)$ in the $x$-$y$ plane.
Note that, for a given momentum, $\phi$ depends on  $\eta$.
In terms of the $\alpha$s and $\beta$s, the Hamiltonian becomes
\begin{align}
	\mathcal{H}
	&=
	\sum_\mathbf{p} \left( \epsilon_- (\mathbf{p}) \alpha^\dagger_\mathbf{p} \alpha_\mathbf{p} + \epsilon_+ (\mathbf{p}) \beta^\dagger_\mathbf{p} \beta_\mathbf{p} \right)
	+\mathcal{H}_{\mathrm{int}}.
\end{align}
The single-particle spectrum has two branches
\begin{align}
	\epsilon_\pm (\mathbf{p}) = \frac{1}{2m}\left[\left(\sqrt{p_x^2 + \eta^2 p_y^2} \pm \kappa\right)^2 + (1 - \eta^2)p_y^2 + p_z^2\right];
	\label{dispersion}
\end{align}
the single particle ground state is given by the lower branch
 $\epsilon_- (\mathbf{p})$, which has degenerate states on the circle $\sqrt{p_x^2 + p_y^2} = \kappa$ and $p_z=0$ when $\eta = 1$, and twofold degeneracy for $\mathbf{p} = (\pm \kappa, 0, 0)$ when $0 \le \eta < 1$.

The interaction part $\mathcal{H}_\mathrm{int}$ is as in Ref.~\cite{Ozawa2011}, except for the $\eta$ dependence:
\begin{align}
	&\mathcal{H}_{\mathrm{int}}
	=
	\frac{1}{V}
	\sum_{\mathbf{p}_1 + \mathbf{p}_2 = \mathbf{p}_3 + \mathbf{p}_4}
	\notag \\
	&
	\left[
	\mathcal{V}^{(1)}_{\phi_1, \phi_2; \phi_3, \phi_4}
	\left(
	\alpha_{\mathbf{p}_4}^\dagger \alpha_{\mathbf{p}_3}^\dagger \alpha_{\mathbf{p}_2} \alpha_{\mathbf{p}_1}
	+
	\beta_{\mathbf{p}_4}^\dagger \beta_{\mathbf{p}_3}^\dagger \beta_{\mathbf{p}_2} \beta_{\mathbf{p}_1}
	\right)/2
	\right.
	\notag \\
	&
	+
	\mathcal{V}^{(2)}_{\phi_1, \phi_2; \phi_3, \phi_4}
	\left(
	\beta_{\mathbf{p}_4}^\dagger \beta_{\mathbf{p}_3}^\dagger \alpha_{\mathbf{p}_2} \alpha_{\mathbf{p}_1}
	+
	\alpha_{\mathbf{p}_4}^\dagger \alpha_{\mathbf{p}_3}^\dagger \beta_{\mathbf{p}_2} \beta_{\mathbf{p}_1}
	\right)/2
	\notag\\
	&
	+
	\mathcal{V}^{(3)}_{\phi_1, \phi_2; \phi_3, \phi_4}
	\left(
	\alpha_{\mathbf{p}_4}^\dagger \beta_{\mathbf{p}_3}^\dagger \beta_{\mathbf{p}_2} \beta_{\mathbf{p}_1}
	+
	\beta_{\mathbf{p}_4}^\dagger \alpha_{\mathbf{p}_3}^\dagger \alpha_{\mathbf{p}_2} \alpha_{\mathbf{p}_1}
	\right)/\sqrt{2}
	\notag \\
	&
	+\mathcal{V}^{(4)}_{\phi_1, \phi_2; \phi_3, \phi_4}
	\left(
	\alpha_{\mathbf{p}_4}^\dagger \alpha_{\mathbf{p}_3}^\dagger \beta_{\mathbf{p}_2} \alpha_{\mathbf{p}_1}
	+
	\beta_{\mathbf{p}_4}^\dagger \beta_{\mathbf{p}_3}^\dagger \alpha_{\mathbf{p}_2} \beta_{\mathbf{p}_1}
	\right)/\sqrt{2}
	\notag \\
	&\left.+
	\mathcal{V}^{(5)}_{\phi_1, \phi_2; \phi_3, \phi_4}
	\alpha_{\mathbf{p}_4}^\dagger \beta_{\mathbf{p}_3}^\dagger \beta_{\mathbf{p}_2} \alpha_{\mathbf{p}_1}
	\right],
\end{align}
where $\phi_i$ is the angle of $(p_{i,x}, \eta p_{i,y})$ in the x-y plane, dependent on $\eta$, and the
$\mathcal{V}^{(i)}$'s are
\begin{align}
	\mathcal{V}^{(1)}_{\phi_1, \phi_2; \phi_3, \phi_4}
	&=
	A_+ +
	\frac{g_{ab}}{8}( e^{i\phi_1} + e^{i\phi_2} ) ( e^{-i\phi_3} + e^{-i\phi_4} )
	\notag\\
	\mathcal{V}^{(2)}_{\phi_1, \phi_2; \phi_3, \phi_4}
	&=A_+ -
	\frac{g_{ab}}{8}( e^{i\phi_1} + e^{i\phi_2} ) ( e^{-i\phi_3} + e^{-i\phi_4} )
	\notag\\
	\mathcal{V}^{(3)}_{\phi_1, \phi_2; \phi_3, \phi_4}
	&=
	\sqrt{2}A_-  +
	\frac{g_{ab}}{4\sqrt{2}}( e^{i\phi_1} + e^{i\phi_2}) ( e^{-i\phi_3} - e^{-i\phi_4})
	\notag\\
	\mathcal{V}^{(4)}_{\phi_1, \phi_2; \phi_3, \phi_4}
	&=
	\sqrt{2}A_- +
	\frac{g_{ab}}{4\sqrt{2}}( e^{i\phi_1} - e^{i\phi_2} ) ( e^{-i\phi_3} + e^{-i\phi_4} )
	\notag\\
	\mathcal{V}^{(5)}_{\phi_1, \phi_2; \phi_3, \phi_4}
	&= 2A_+ -
	\frac{g_{ab}}{4}( e^{i\phi_1} - e^{i\phi_2} ) ( e^{-i\phi_3} - e^{-i\phi_4}), \label{vertices}
\end{align}
with
\begin{align}
  A_\pm &= (g_{aa} \pm g_{bb} e^{i(\phi_1+\phi_2-\phi_3-\phi_4)})/4.
\end{align}

\section{the $t$-matrix}

We now turn to calculating the scattering $t$ matrix,  $\Gamma_{\alpha \alpha}^{\alpha \alpha} (\mathbf{p}, \mathbf{p}^\prime; \mathbf{q})$, for 
 two bosons in the ground states of the  lower ($\alpha$) branch, with incoming momenta $\mathbf{q}/2 + \mathbf{p}$ and $\mathbf{q}/2 - \mathbf{p}$, and outgoing momenta $\mathbf{q}/2 + \mathbf{p}^\prime$ and $\mathbf{q}/2 - \mathbf{p}^\prime$.  Note that, for particles in the single particle ground state, the $z$ component of the momentum is zero. 
As we showed earlier, the ultraviolet divergences in the $t$ matrix can be renormalized in terms of cutoff-independent low-energy parameters \cite{Ozawa2011}, and, in the isotropic case, the $t$ matrix depends only on the spin-orbit coupling strength, the scattering lengths, and the incident and outgoing momenta.
 The exact summation of ladder diagrams for the $t$ matrix, for both isotropic and anisotropic spin-orbit couplings, is then (see Appendix A for the derivation):
\begin{align}
	&\Gamma_{\alpha \alpha}^{\alpha \alpha} (\mathbf{p}, \mathbf{p}^\prime; \mathbf{q})
	=
	\notag \\
	&\frac{\pi}{m\kappa}
	\begin{pmatrix}
	1 & e^{i(\phi_1 + \phi_2)} & \dfrac{e^{i\phi_1} + e^{i\phi_2}}{2}
	\end{pmatrix}
	M^{-1}
	\begin{pmatrix}
	1 \\ e^{-i(\phi_3 + \phi_4)} \\ \dfrac{e^{-i\phi_3} + e^{-i\phi_4}}{2}
	\end{pmatrix},
	\label{exactt}
\end{align}
where
\begin{align}
	&M
	=
	\notag \\
	&\begin{pmatrix}
	f(\frac{\tilde{q}}{2}) + \frac{1}{\kappa a_{aa}} & h_1 (\tilde{\mathbf{q}}/2) & h_2 (\tilde{\mathbf{q}}/2) \\
	h_1^* (\tilde{\mathbf{q}}/2) & f(\frac{\tilde{q}}{2}) + \frac{1}{\kappa a_{bb}} & h_2^* (\tilde{\mathbf{q}}/2) \\
	h_2^* (\tilde{\mathbf{q}}/2) & h_2 (\tilde{\mathbf{q}}/2) & \frac{1}{2}\left(f(\frac{\tilde{q}}{2}) - g(\frac{\tilde{q}}{2}) + \frac{1}{\kappa a_{ab}} \right)
	\end{pmatrix},
\end{align}
with ${\bar {\mathbf q}} \equiv {\mathbf q}/\kappa$ and $\tilde{q} \equiv q/\kappa$.
The angles $\phi_1$, $\phi_2$, $\phi_3$, and $\phi_4$ are those of the vectors $\mathbf{q}/2+\mathbf{p}$, $\mathbf{q}/2-\mathbf{p}$, $\mathbf{q}/2-\mathbf{p}^\prime$, and $\mathbf{q}/2+\mathbf{p}^\prime$ in the $x$-$y$ plane with a factor of $\eta$ multiplying the $y$ components.  
The dimensionless functions $f(\tilde{q}/2)$, $g(\tilde{q}/2)$, $h_1 (\tilde{\mathbf{q}}/2)$, and $h_2 (\tilde{\mathbf{q}}/2)$ are,
\begin{align}
	&f (\tilde{q}/2)
	\equiv
	\frac{\pi}{m\kappa}\int \frac{d^3 k}{(2\pi)^3}
	\left[
	\frac{1}{\epsilon_- (\frac{\mathbf{q}}{2} + \mathbf{k}) + \epsilon_- (\frac{\mathbf{q}}{2} - \mathbf{k})}
	+
	\right.
	\notag \\
	&\left.
	\frac{1}{\epsilon_+ (\frac{\mathbf{q}}{2} + \mathbf{k}) + \epsilon_+ (\frac{\mathbf{q}}{2} - \mathbf{k})}
	+\frac{2}{\epsilon_- (\frac{\mathbf{q}}{2} + \mathbf{k}) + \epsilon_+ (\frac{\mathbf{q}}{2} - \mathbf{k})}
	-
	\frac{4m}{k^2}
	\right] \notag \\
	&g (\tilde{q}/2)
	\equiv
	-\frac{\pi}{m\kappa} \int \frac{d^3 k}{(2\pi)^3}
	\left[
	\frac{\cos (\phi_5 - \phi_6)}{\epsilon_- (\frac{\mathbf{q}}{2} + \mathbf{k}) + \epsilon_- (\frac{\mathbf{q}}{2} - \mathbf{k})}
	\right.
	\notag \\
	&\left.+
	\frac{\cos (\phi_5 - \phi_6)}{\epsilon_+ (\frac{\mathbf{q}}{2} + \mathbf{k}) + \epsilon_+ (\frac{\mathbf{q}}{2} - \mathbf{k})}
	-\frac{2\cos (\phi_5 - \phi_6)}{\epsilon_- (\frac{\mathbf{q}}{2} + \mathbf{k}) + \epsilon_+ (\frac{\mathbf{q}}{2} - \mathbf{k})}
	\right]
	\notag \\
	&h_1 (\tilde{\mathbf{q}}/2)
	\equiv
	\frac{\pi}{m\kappa}\int \frac{d^3 k}{(2\pi)^3}
	\left[
	\frac{e^{i(\phi_5 + \phi_6)}}{\epsilon_- (\frac{\mathbf{q}}{2} + \mathbf{k}) + \epsilon_- (\frac{\mathbf{q}}{2} - \mathbf{k})}
	\right.
	\notag \\
	&\left.+
	\frac{e^{i(\phi_5 + \phi_6)}}{\epsilon_+ (\frac{\mathbf{q}}{2} + \mathbf{k}) + \epsilon_+ (\frac{\mathbf{q}}{2} - \mathbf{k})}
	-
	\frac{2e^{i(\phi_5 + \phi_6)}}{\epsilon_- (\frac{\mathbf{q}}{2} + \mathbf{k}) + \epsilon_+ (\frac{\mathbf{q}}{2} - \mathbf{k})}
	\right]
	\notag \\
	&h_2 (\tilde{\mathbf{q}}/2)
	\equiv
	\frac{\pi}{2m\kappa}\int \frac{d^3 k}{(2\pi)^3}
	\left[
	\frac{e^{i \phi_5} + e^{i \phi_6}}{\epsilon_- (\frac{\mathbf{q}}{2} + \mathbf{k}) + \epsilon_- (\frac{\mathbf{q}}{2} - \mathbf{k})}
	\right.
	\notag \\
	&\left.-
	\frac{e^{i \phi_5} + e^{i \phi_6}}{\epsilon_+ (\frac{\mathbf{q}}{2} + \mathbf{k}) + \epsilon_+ (\frac{\mathbf{q}}{2} - \mathbf{k})}
	-
	\frac{2(e^{i \phi_5} - e^{i \phi_6})}{\epsilon_- (\frac{\mathbf{q}}{2} + \mathbf{k}) + \epsilon_+ (\frac{\mathbf{q}}{2} - \mathbf{k})}
	\right],
\end{align}
where $\phi_5$ and $\phi_6$ are the angles of $\mathbf{q}/2 - \mathbf{k}$ and $\mathbf{q}/2+\mathbf{k}$ in the $x$-$y$ plane with $y$ components multiplied by $\eta$.
Changing the angle of $\mathbf{q}$ in the x-y plane only changes the overall phases of $h_1 (\tilde{\mathbf{q}}/2)$ and $h_2 (\tilde{\mathbf{q}}/2)$.   These four functions are everywhere finite except for 
the logarithmic divergence of $f(\tilde{q}/2)$ at $\tilde{q} = 0$.
When the scattering lengths are small, the diagonal elements of $M$ are dominant and we may ignore the off-diagonal elements; the $t$ matrix thus obtained does not have terms containing products of different scattering lengths, which is
the approximate result obtained for the bosonic $t$ matrix in \cite{Ozawa2011}.

\section{Ground state phases}

We now determine the many-body ground state via mean-field theory using the $t$ matrix derived above as the effective interactions, an approximation valid as long as the $na_{ij}^3$ are all  $\ll 1$, where $n$ is the particle density.
In mean field, we assume that all particles are in the single-particle ground states $(\kappa, 0, 0)$ or $(-\kappa, 0, 0)$, and thus ignore possible occupation of excited states as a consequence of the interaction.  In this case the system is described essentially by the Nozi\`eres model \cite{Nozieres1995}.  
The issues of going beyond mean field, (e.g., via Bogoliubov theory) as well as including possible effects of the condensate on the effective interaction, are beyond the scope of this paper and are left for the future.
For $0 \le \eta < 1$, 
we take the particles to be either at $\mathbf{p} = (\kappa, 0, 0)$ or $(-\kappa, 0, 0)$;
the relevant interactions are those between particles of either the same momentum or opposite momenta.
We denote the interaction with same momentum by $\Gamma_0 \equiv \Gamma_{\alpha \alpha}^{\alpha \alpha}(0,0;\pm 2{\mathbf K})$ and that with opposite momenta by $\Gamma_\pi \equiv \Gamma_{\alpha \alpha}^{\alpha \alpha}(\pm 
\mathbf{K},\mp \mathbf{K},0)$, where we use the abbreviated notation $\mathbf{K} \equiv (\kappa, 0,0)$. 

The relevant terms in the interaction are then equivalent to the Nozi\`eres model \cite{Nozieres1995}
\begin{align}
	&\mathcal{H}_{\mathrm{int}}
	\sim
	\frac{1}{2V} \Gamma_0 N(N+1)\
	+ \frac{1}{V}(2\Gamma_\pi - \Gamma_0) N_\pi N_0,
	\label{nozieres}
\end{align}
where $N_0 \equiv \alpha^\dagger_{(\kappa, 0, 0)} \alpha_{(\kappa, 0, 0)}$ and $N_\pi \equiv \alpha^\dagger_{(-\kappa, 0, 0)} \alpha_{(-\kappa, 0, 0)}$.
The total number of particles, $N = N_0 + N_\pi$, is fixed.
For $\Gamma_0 < 2\Gamma_\pi$,
the ground state is a single BEC with either all the particles in $(\kappa, 0, 0)$ or $(-\kappa, 0, 0)$, while, for
$\Gamma_0 > 2\Gamma_\pi$, the condensate is nominally fragmented with half of the atoms forming a BEC in one state and the other half forming a BEC in the other state.
However, as shown in Ref.~\cite{Mueller2006}, such a fragmented state is expected to be unstable against formation of a coherent condensate with a condensate wave function that is a coherent superposition of the two momenta.
Following the conventions of Refs.~\cite{Wang2010,Jian2011}, we call the single BEC phase ``plane wave," and the BEC phase with two different momenta ``striped."
The difference of the present calculation from earlier studies with mean-field couplings \cite{Wang2010,Jian2011}, is that here the bare couplings, $\mathcal{V}^{(1)}_{0,0;0,0}$ and $\mathcal{V}^{(1)}_{0,\pi;0,\pi}$, are replaced by $\Gamma_0$ and $\Gamma_\pi$, respectively.

\begin{figure}[htbp]
\begin{center}
\subfigure[]{
\includegraphics[scale=0.28]{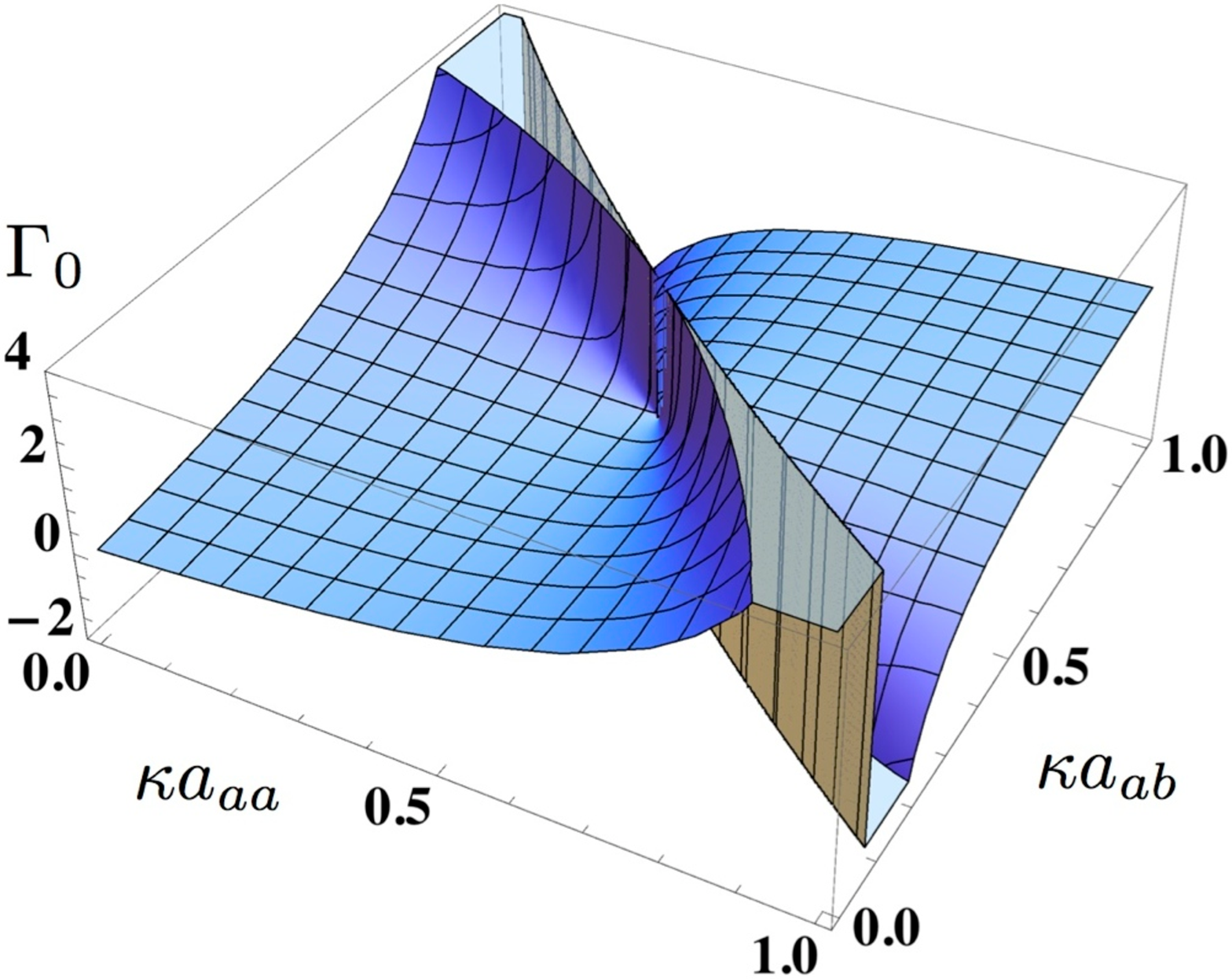}}
\subfigure[]{
\includegraphics[scale=0.8]{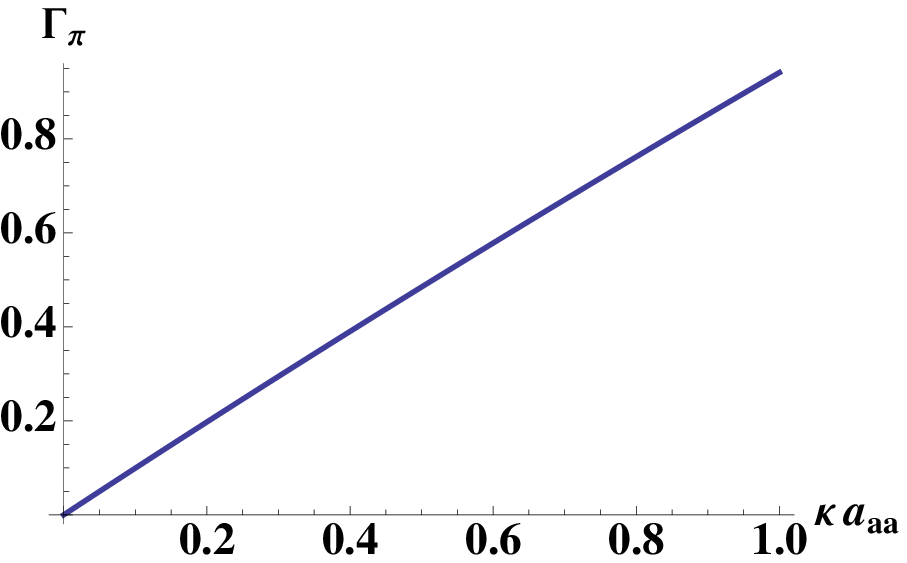}}
\caption{(Color online) (a) $\Gamma_0$ as a function of $\kappa a_{aa}$ and $\kappa a_{ab}$, and (b) $\Gamma_\pi$ as a function of $\kappa a_{aa}$, both scaled by $2\pi/(m\kappa)$, for $\eta = 0.5$. The vertical plane in the middle of panel (a) indicates the resonance where, from left to right, $\Gamma_0$ diverges to positive infinity and comes back from negative infinity.}
\label{gamma0etpieta05}
\end{center}
\end{figure}

While there is no difficulty in deriving the phase diagrams for general scattering lengths, we assume here for simplicity that the intraspecies scattering lengths are equal, $a_{aa} = a_{bb}$.  Then, 
\begin{align}
	&\Gamma_0
	=
	\frac{2\pi}{m\kappa}\times
	\notag \\
	&
	\frac{1/\kappa a_{aa} + 1/\kappa a_{ab} + 2f(1) - g(1) +h_1(1)-4h_2(1)}{(1/\kappa a_{ab} + f(1)-g(1))(1/\kappa a_{aa} + f(1) +h_1(1))-4h_2(1)^2}
	\notag \\
	&\Gamma_\pi
	=
	\frac{2\pi}{m\kappa} \frac{1}{1/\kappa a_{aa} + f(0) - h_1 (0)}, \label{gamma0pi}
\end{align}
where $h_1(0) \equiv h_1 (\tilde{\mathbf{q}}=(0,0,0))$, $h_1(1) \equiv h_1 (\tilde{\mathbf{q}}=(1,0,0))$, etc.
The quantities $f(0), h_1(0), f(1), g(1), h_1(1)$, and $h_2(1)$, which depend on $\eta$, can
be calculated numerically.
The interaction between different momenta $\Gamma_\pi$ is independent of $a_{ab}$,
and is a monotonically increasing nonnegative function of $\kappa a_{aa} $, equal to $0$ at $\kappa a_{aa} = 0$ and reaching $2\pi/\left[m\kappa(f(0)-h_1(0))\right]$ at $\kappa a_{aa} = \infty$.
The dependence of $\Gamma_0$ on $\kappa a_{aa}$ and $\kappa a_{ab}$ is more complicated.
We plot $\Gamma_0$ and $\Gamma_\pi$, both scaled by $2\pi/(m\kappa)$, for $\eta = 0.5$ in Fig. \ref{gamma0etpieta05}.
Now we discuss the ground-state phases from $\eta = 0$ to 1.

When $\eta = 0$, the effective interactions are relatively simple.
It can be shown that $f(0) = h_1 (0)$ for $\eta = 0$; hence $\Gamma_\pi = 2\pi a_{aa}/m$, and the effective interaction in the $\mathbf{q} = 0$ channel does not depend on the spin-orbit coupling strength $\kappa$.
In the $\mathbf{q}/2 = (\kappa, 0, 0)$ channel,
$f(1) = -1, g(1) = 0, h_1(1) = 0$, and $h_2(1) = 1/2$,
so,
$$\displaystyle \Gamma_0 = \frac{2\pi}{m\kappa}\frac{\kappa a_{aa} + \kappa a_{ab} - 4 \kappa a_{aa} \kappa a_{ab}}{1 - \kappa a_{aa} - \kappa a_{ab}}$$
for $\eta = 0$.
The effective interaction at small $\kappa a_{aa}$ and $\kappa a_{ab}$ is positive,
and diverges when $\kappa a_{aa} + \kappa a_{ab}$ approaches unity.
As one crosses the line $\kappa a_{aa} + \kappa a_{ab} = 1$,
$\Gamma_0$ starts at negative infinity and remains negative until $\kappa a_{aa} + \kappa a_{ab} = 4\kappa a_{aa} \kappa a_{ab}$, after which $\Gamma_0$ is positive.
When $\Gamma_0$ is negative, we expect the BEC  in bulk to be unstable against collapse,
as in ordinary BECs with negative scattering length in the absence of spin-orbit couplings.   We call the phase with an attractive interaction ``unstable."
The three possible ground-state phases,  plane wave, striped, and unstable, are determined by the sign of $\Gamma_0$ and the interplay between $\Gamma_0$ and $\Gamma_\pi$.

As $\eta$ increases from 0, the basic structure of $\Gamma_0$ does not change;
$\Gamma_0$ remains positive at small $\kappa a_{aa}$ and $\kappa a_{ab}$, and
as these variables increase, $\Gamma_0$ again diverges at a line in the $\kappa a_{aa}$-$\kappa a_{ab}$ plane,
beyond which it is 
negative up to a second line, after which $\Gamma_0$ is positive.   Since the denominator of Eq.~(\ref{gamma0pi}) for
$\Gamma_0$ is quadratic in $1/(\kappa a)$, it has in fact two zeros, the one for positive scattering lengths, as shown, and a second for negative scattering lengths, which is discussed at the end of this section.  The structure for positive scattering lengths is illustrated in  Fig.~\ref{gamma0etpieta05}, for $\eta = 0.5$.  

\begin{figure}[htbp]
\begin{center}
\subfigure[$\eta = 0$]{
\includegraphics[scale=0.453]{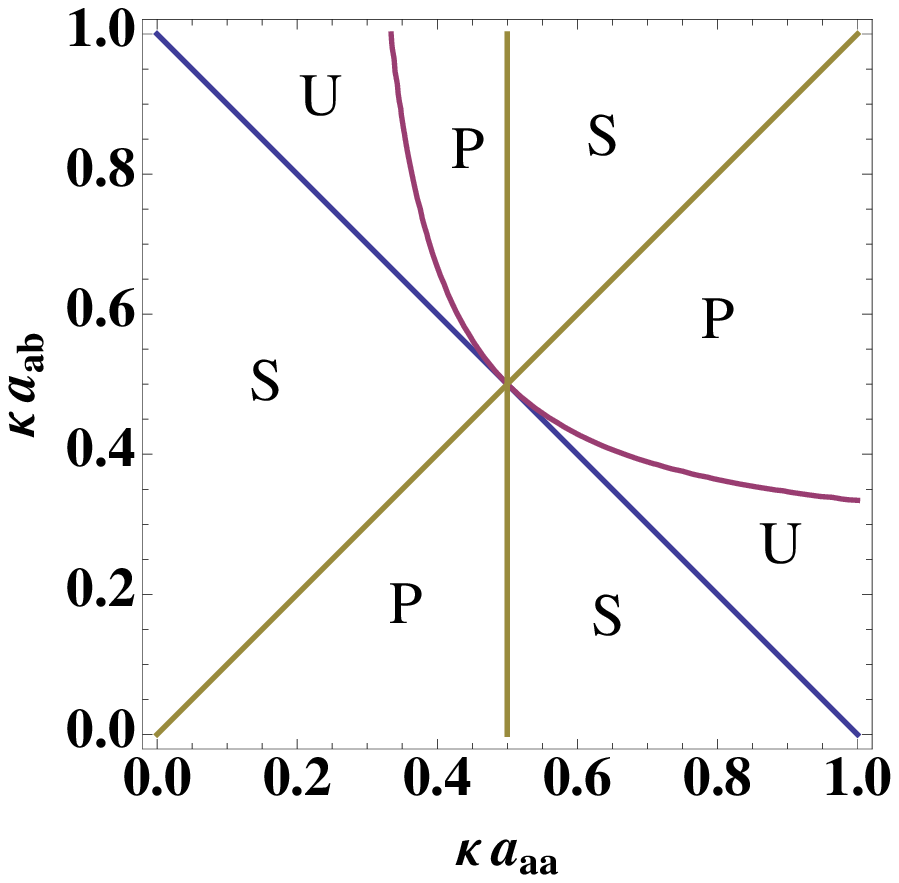}}
\subfigure[$\eta = 0.25$]{
\includegraphics[scale=0.453]{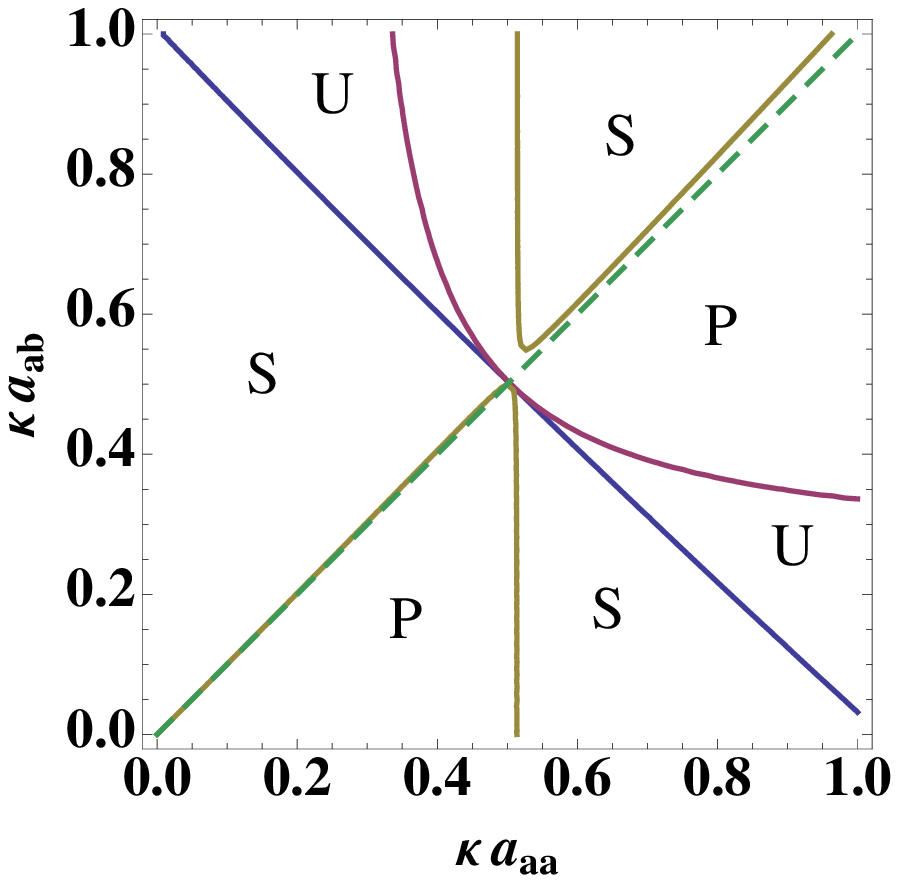}}
\subfigure[$\eta = 0.50$]{
\includegraphics[scale=0.453]{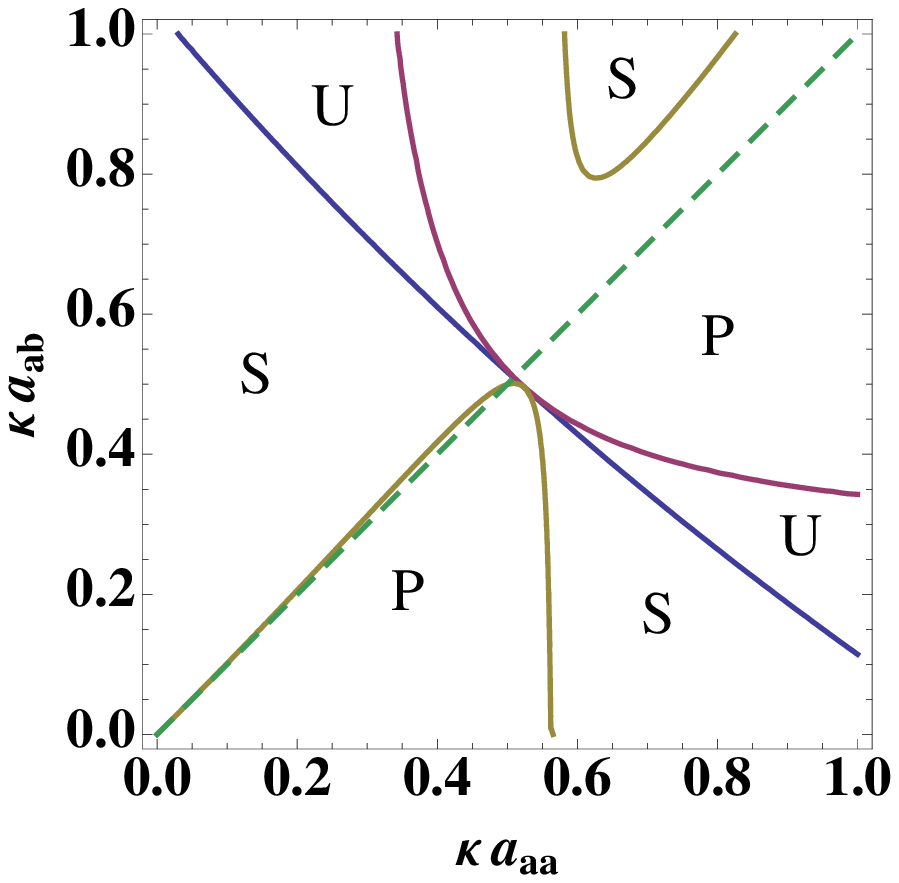}}
\subfigure[$\eta = 0.75$]{
\includegraphics[scale=0.453]{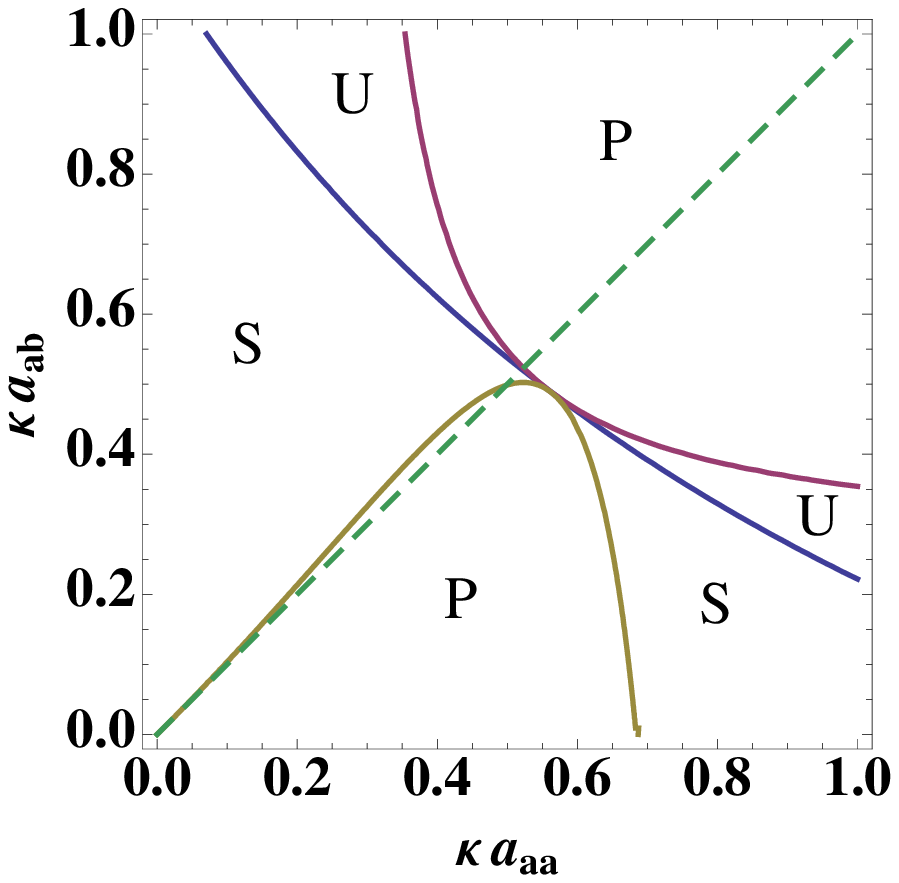}}
\caption{(Color online) Ground state phase diagrams in the $\kappa a_{aa}$ - $\kappa a_{ab}$ plane for anisotropies $\eta = 0$, 0.25, 0.5, and 0.75. The regions P are the plane-wave phase with a BEC of a single momentum. The regions S are the striped phase with a BEC of a coherent superposition of two different momenta. The phase in the regions U are unstable, with  the effective interaction $\Gamma_0$  negative.   Along the line between S and U, $\Gamma_0$ diverges, and along the line between U and P, $\Gamma_0$ vanishes.  The intersection of these two lines is a critical point.  The dashed lines indicate the phase diagram derived using mean-field coupling, in which the plane is separated into
an upper striped region and a lower plane wave region.}
\label{zeroto075}
\end{center}
\end{figure}

The ground-state phase diagrams for various $\eta$ are plotted in Fig.~\ref{zeroto075}. 
In the panels, the plane-wave phase is labeled ``P," the striped phase ``S," and the unstable phase ``U."
The plane-wave phase occurs when $\Gamma_0 < 2\Gamma_\pi$, the striped phase when $\Gamma_0 > 2\Gamma_\pi$, and the unstable phase when $\Gamma_0 < 0$.
Note the overall tendency of the phase diagrams as $\eta$ increases;
the upper striped region detaches from the resonant critical point, where the resonant line (between S and U) and the line with $\Gamma_0 = 0$ (between U and P) touch, and the region is pushed upward as $\eta$ is increased. Meanwhile, the shapes of the resonant line and the boundaries of plane-wave regions change but,
with the exception of the upper striped region,
 the overall topology does not change.
The dashed lines $a_{aa} = a_{ab}$ in the figures are the phase-separation lines obtained earlier \cite{Wang2010} using mean-field couplings  $4\pi a_{aa}/m$ and $4\pi a_{ab}/m$; there the striped phase is preferred above the dashed lines and the plane-wave phase is preferred below the dashed lines.
Use of mean-field couplings is accurate for small $\kappa a_{aa}$ and $\kappa a_{ab}$ but, as these variables increase, the deviation from the mean-field-coupling prediction becomes significant
and the phase diagrams exhibit qualitatively new and rich structures.

\begin{figure}[htbp]
\begin{center}
\subfigure[$\eta = 0.999$]{
\includegraphics[scale=0.453]{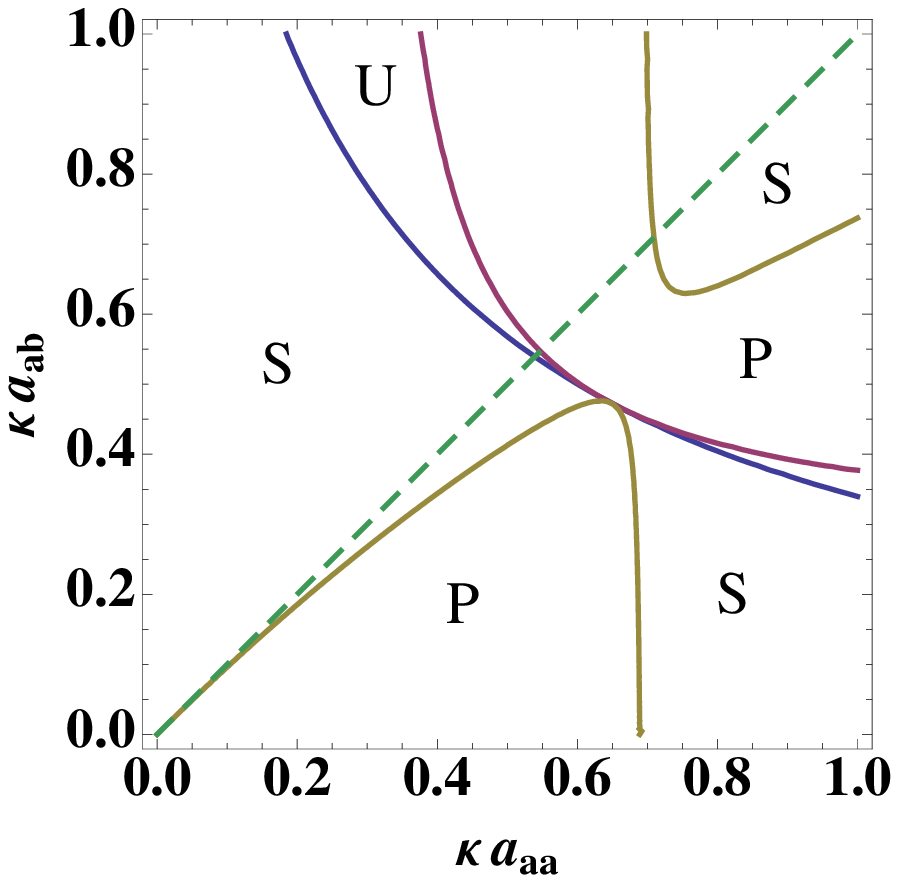}}
\subfigure[$\eta = 0.9995$]{
\includegraphics[scale=0.453]{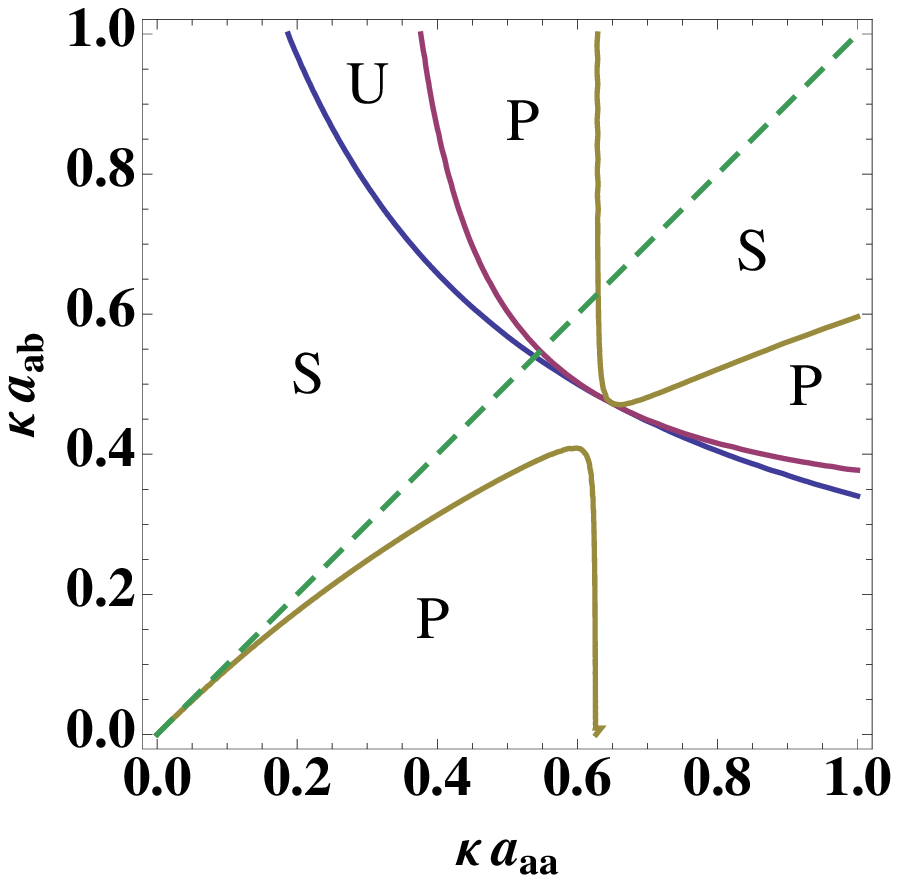}}
\subfigure[$\eta = 0.99999$]{
\includegraphics[scale=0.453]{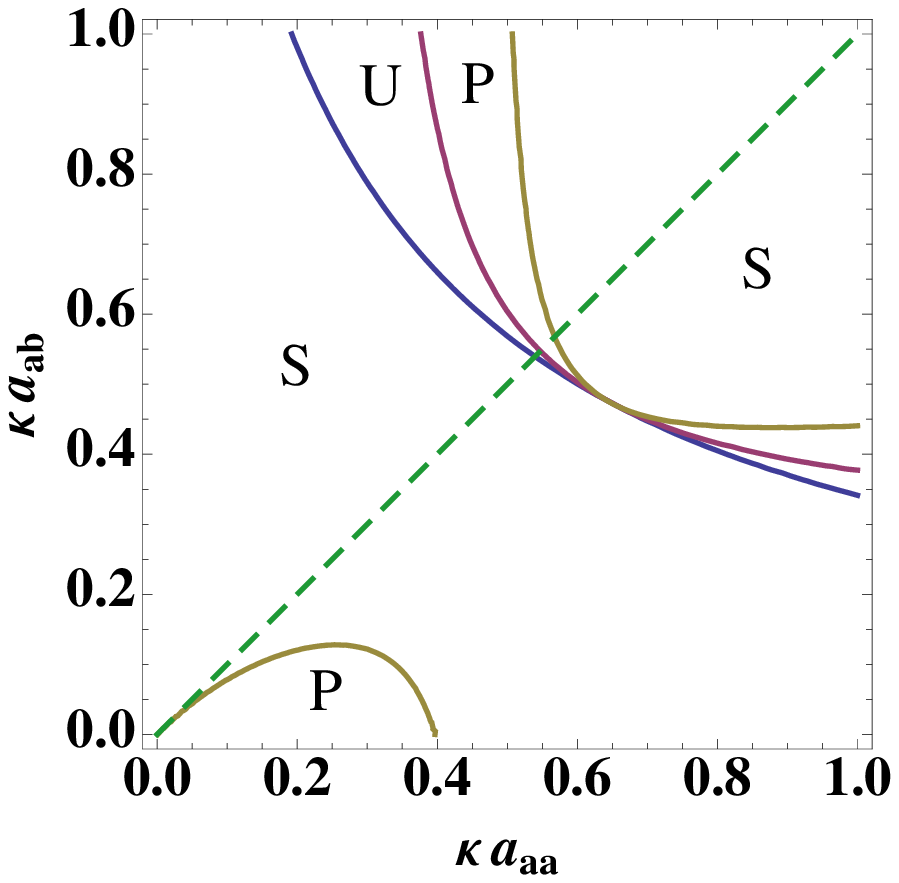}}
\subfigure[$\eta = 1$]{
\includegraphics[scale=0.453]{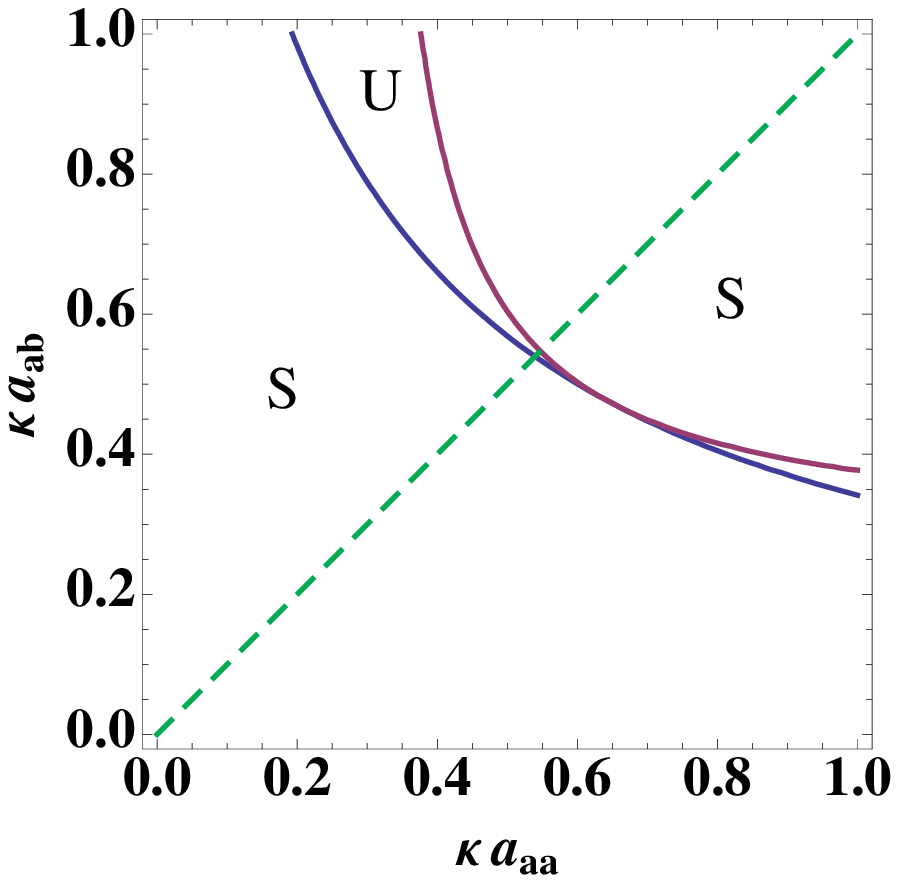}}
\caption{(Color online) Ground-state phase diagrams for $\eta$ close to unity.}
\label{zero999to1}
\end{center}
\end{figure}

This overall tendency continues to around $\eta \sim 0.99$.
With further increase of $\eta$ toward isotropy, $\eta = 1$,
we start to observe qualitatively new behavior of the phase diagrams.
The phase diagrams close to $\eta = 1$ are plotted in Fig.~\ref{zero999to1}.
As one sees, the striped region comes back from above and touches the resonant critical point, 
and at the same time the lower plane-wave region detaches from the critical point.
In the limit $\eta = 1$, the plane-wave region vanishes.

The behavior around $\eta \simeq 1$ is in fact logarithmic in the deviations of the anisotropy $\eta$ from unity.  We write $\delta = 1 - \eta^2$;
as $\delta \to 0$, $h_1(0)$, $f(1)$, $g(1)$, $h_1(1)$, and $h_2 (1)$ approach finite values,
but, in leading order for small $\delta$, $f(0) \sim |\ln \delta |/4$.  Setting, for small $\delta$, $h_1(0)$, $f(1)$, $g(1)$, $h_1(1)$, and $h_2 (1)$, to their values at $\delta = 0$ and approximating $f(0)$ by $|\ln \delta |/4$,
corresponds to fixing $\Gamma_0$ and varying the slope of $\Gamma_\pi$.
In the isotropic limit $\delta = 0$, $\Gamma_\pi = 0$ and thus a plane-wave region is not allowed [cf. Eq.~(\ref{nozieres})].
With small anisotropy, $\Gamma_\pi$ can be positive, and small plane-wave regions appear.

\begin{figure}[htbp]
\begin{center}
\includegraphics[width=6.5cm]{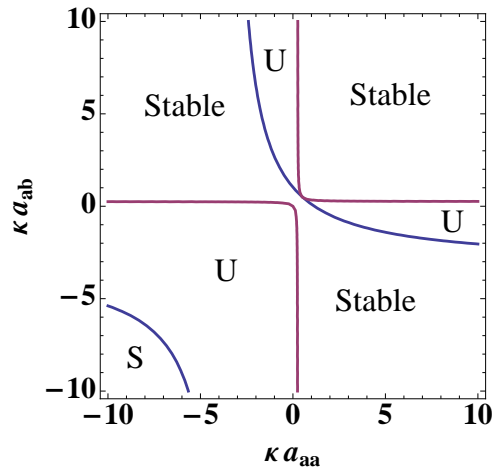}
\caption{(Color online) Ground-state phase diagram for $\eta = 0.5$ extended to negative values of scattering lengths. The regions marked U and S are unstable and striped phases, as before. The region marked ``Stable" is either a plane-wave or striped phase.  Note the appearance of a stable (striped) phase when both scattering lengths are large and negative.}
\label{negativeeta05}
\end{center}
\end{figure}

We briefly consider tuning the scattering lengths to negative values.
In the absence of spin-orbit couplings, negative scattering lengths lead to an instability in large systems.
On the other hand, as we see from Eq.~(\ref{gamma0pi}), tuning the inverse scattering lengths to just below $0$ does not immediately lead to an attractive interaction; in the presence of the spin-orbit coupling fields, Rashba-Dresselhaus couplings can stabilize BECs with negative scattering lengths if the inverse scattering lengths are small.  Even when $\Gamma_0$ is negative, systems with small particle number can be metastable in the presence of an attractive interaction\footnote{
 Assuming bosons trapped in an isotropic harmonic potential, we can roughly estimate the particle number below which the condensate is stable with $\Gamma_0<0$.
In the absence of spin-orbit coupling, the critical number of bosons is $N_c \sim 0.6 a_{\mathrm{osc}}/|a|$, where $a_{\mathrm{osc}}$ is the oscillator length of a trap $\sqrt{\hbar/(m\omega)}$ \cite{Baym1996}.
For spin-orbit coupled bosons, the scattering length is replaced by $m\Gamma_0/(4\pi)$.
Introducing a scaled effective coupling $\tilde{\Gamma}_0 = m\kappa \Gamma_0/(2\pi)$ (the scale used in
Fig.~\ref{gamma0etpieta05}), we estimate a critical number $N_c \sim 0.3 \kappa a_{\mathrm{osc}}/|\tilde{\Gamma}_0|$.
Using realistic values of $\kappa \sim \sqrt{2}\pi/804$ nm \cite{Lin2011} and $a_{\mathrm{osc}} \sim 1\mu$m, we obtain $N_c \sim 2/|\tilde{\Gamma}_0|$, which implies that stabilization occurs only quite close to the line $\Gamma_0 = 0$.}.
For illustration, we plot the phase diagram for $\eta = 0.5$ extended to negative scattering lengths in Fig.~\ref{negativeeta05}.
In the regions marked ``Stable,"  $\Gamma_0 > 0$ and the ground state is either a plane-wave or striped phase.  As seen in the figure, when both scattering lengths $a_{aa}$ and $a_{ab}$ are negative and large, another stable region  appears in the phase diagram, in which the ground state is in the striped phase.  The line between the lower-left striped phase and the unstable phase is a second resonant line along which $\Gamma_0$ diverges.
A stable region with negative scattering lengths generally exists for all $0 < \eta \le 1$; as $\eta$ increases, the stable region in the phase diagram becomes larger.

\section{Conclusion}

\begin{figure}[htbp]
\begin{center}
\subfigure[]{
\includegraphics[scale=0.6]{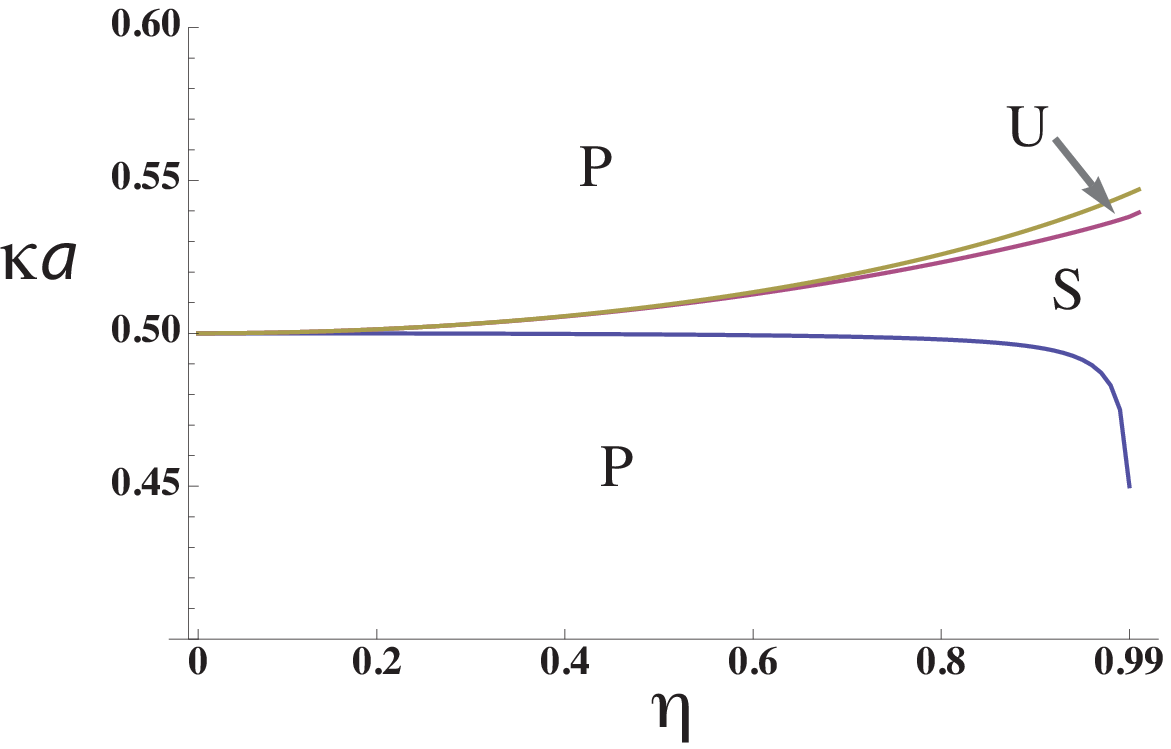}}
\subfigure[]{
\includegraphics[scale=0.73]{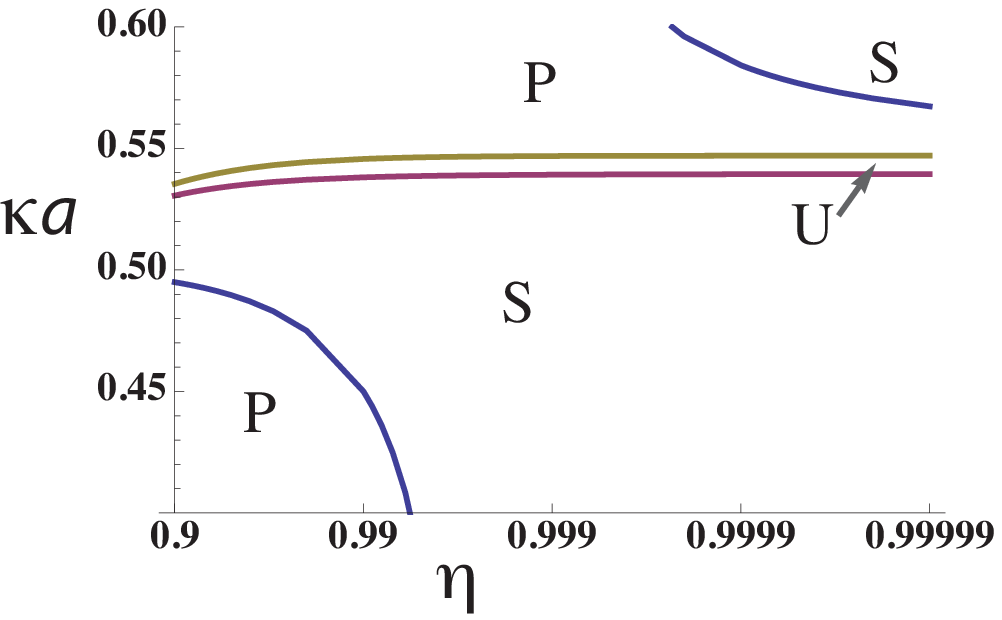}}
\caption{(Color online) Ground state phase diagram when $a_{aa} = a_{bb} = a_{ab} = a$ in the $\eta$-$\kappa a$ plane for (a) anisotropies less than 0.99 and (b) anisotropies close to unity. The horizontal axis of panel (b) is a logarithmic scale.}
\label{su2_phases}
\end{center}
\end{figure}

Proposed schemes to realize Rashba-Dresselhaus spin-orbit couplings in ultracold atomic experiments \cite{Ruseckas2005,Zhu2006,Liu2009,Juzeliunas2010,Campbell2011} use Raman lasers to couple atoms in different hyperfine states. 
In general as one transforms the original basis to one in which the coupling has the Rashba-Dresselhaus spin-orbit structure, the interaction Hamiltonian acquires terms such as $a^\dagger_{\mathbf{p}_4} a^\dagger_{\mathbf{p}_3} a_{\mathbf{p}_2} b_{\mathbf{p}_1}$ which do not conserve the number of particles in each pseudospin state ($a$-like and $b$-like). 
Our analysis, which did not take such terms into account, can be directly compared with proposed experiments  when the interaction is independent of species ($g_{aa} = g_{bb} = g_{ab}$), in which case the interaction is independent of the choice of basis.
This condition is a good approximation for the three hyperfine states of $^{87}$Rb in the lowest $F = 1$ state.
The assumption that $g_{aa} = g_{bb} = g_{ab}$ corresponds to the (dashed) diagonal lines in Figs.~\ref{zeroto075} and \ref{zero999to1}.
Figure~\ref{su2_phases} shows the phase diagram in the $\eta$-$\kappa a$ plane, where $a$ is the assumed common scattering length.

For $0 \le \eta \le 0.99$ the system, with increasing $\kappa a$, experiences transitions from plane-wave to striped, then to unstable, and finally to the plane-wave phase again, as seen in Fig.~\ref{su2_phases}(a).
Looking more closely at the region $0.9 \le \eta \le 1$, as drawn on a logarithmic scale in Fig.~\ref{su2_phases}(b), we find that the line separating the lower plane-wave and striped regions terminates and another line starts from positive infinity above which the striped phase is preferred. This new line touches the uppermost line (below the upper P phase) in the figure in the $\eta \to 1$ limit and thus no plane-wave region exists at isotropic spin-orbit coupling\footnote{While one can achieve large scattering lengths experimentally with Feshbach resonances, the general $m_F, m_{F^\prime}$ dependence of the resonances leads to differences of the scattering lengths near the resonances, a complicating feature requiring analysis beyond the scope of this paper.}.

\begin{acknowledgements}
We are extremely grateful to Professor Tin-Lun Ho for critical comments.  This research was supported in part by NSF Grant PHY09-69790.
\end{acknowledgements}

\appendix
\section{Exact summation of ladder diagrams for the $t$-matrix}
Here we outline the derivation of the exact $t$ matrix (\ref{exactt}).
Our starting point is the set of Bethe-Salpeter equations:
\begin{align}
	&\Gamma_{\alpha \alpha}^{\alpha \alpha} (\mathbf{p}, \mathbf{p}^\prime; \mathbf{q})
	=
	\mathcal{V}^{(1)}_{\phi_1, \phi_2; \phi_3, \phi_4}
	\notag \\
	&
	-
	\int \frac{d^3 k}{(2\pi)^3}
	\left[
	\frac{\mathcal{V}^{(1)}_{\phi_1, \phi_2; \phi_5, \phi_6} \Gamma_{\alpha \alpha}^{\alpha \alpha}(\mathbf{k}, \mathbf{p}^\prime; \mathbf{q})}{\epsilon_- (\frac{\mathbf{q}}{2} - \mathbf{k}) + \epsilon_- (\frac{\mathbf{q}}{2}+\mathbf{k})}
	\right.
	\notag \\
	&\left.
	+
	\frac{\mathcal{V}^{(2)}_{\phi_1, \phi_2; \phi_5, \phi_6} \Gamma_{\beta \beta}^{\alpha \alpha}(\mathbf{k}, \mathbf{p}^\prime; \mathbf{q})}{\epsilon_+ (\frac{\mathbf{q}}{2} - \mathbf{k}) + \epsilon_+ (\frac{\mathbf{q}}{2}+\mathbf{k})}
	+
	\frac{\mathcal{V}^{(3)}_{\phi_1, \phi_2; \phi_6, \phi_5} \Gamma_{\alpha \beta}^{\alpha \alpha}(\mathbf{k},\mathbf{p}^\prime; \mathbf{q})}{\epsilon_+ (\frac{\mathbf{q}}{2} - \mathbf{k}) + \epsilon_- (\frac{\mathbf{q}}{2}+\mathbf{k})}
	\right]
	\notag
	\\
	&\Gamma_{\beta \beta}^{\alpha \alpha} (\mathbf{p}, \mathbf{p}^\prime; \mathbf{q})
	=
	\mathcal{V}^{(2)}_{\phi_1, \phi_2; \phi_3, \phi_4}
	\notag \\
	&
	-
	\int \frac{d^3 k}{(2\pi)^3}
	\left[
	\frac{\mathcal{V}^{(2)}_{\phi_1, \phi_2; \phi_5, \phi_6} \Gamma_{\alpha \alpha}^{\alpha \alpha}(\mathbf{k}, \mathbf{p}^\prime; \mathbf{q})}{\epsilon_- (\frac{\mathbf{q}}{2} - \mathbf{k}) + \epsilon_- (\frac{\mathbf{q}}{2}+\mathbf{k})}
	\right.
	\notag \\
	&\left.
	+
	\frac{\mathcal{V}^{(1)}_{\phi_1, \phi_2; \phi_5, \phi_6} \Gamma_{\beta \beta}^{\alpha \alpha}(\mathbf{k}, \mathbf{p}^\prime; \mathbf{q})}{\epsilon_+ (\frac{\mathbf{q}}{2} - \mathbf{k}) + \epsilon_+ (\frac{\mathbf{q}}{2}+\mathbf{k})}
	+
	\frac{\mathcal{V}^{(3)}_{\phi_1, \phi_2; \phi_5, \phi_6} \Gamma_{\alpha \beta}^{\alpha \alpha}(\mathbf{k},\mathbf{p}^\prime; \mathbf{q})}{\epsilon_+ (\frac{\mathbf{q}}{2} - \mathbf{k}) + \epsilon_- (\frac{\mathbf{q}}{2}+\mathbf{k})}
	\right]
	\notag 
	\\
	\rm{and}  \nonumber\\
	&\Gamma_{\alpha \beta}^{\alpha \alpha} (\mathbf{p}, \mathbf{p}^\prime; \mathbf{q})
	=
	\mathcal{V}^{(4)}_{\phi_1, \phi_2; \phi_3, \phi_4}
	\notag \\
	&
	-
	\int \frac{d^3 k}{(2\pi)^3}
	\left[
	\frac{\mathcal{V}^{(4)}_{\phi_1, \phi_2; \phi_5, \phi_6} \Gamma_{\alpha \alpha}^{\alpha \alpha}(\mathbf{k}, \mathbf{p}^\prime; \mathbf{q})}{\epsilon_- (\frac{\mathbf{q}}{2} - \mathbf{k}) + \epsilon_- (\frac{\mathbf{q}}{2}+\mathbf{k})}
	\right.
	\notag \\
	&\left.
	+
	\frac{\mathcal{V}^{(4)}_{\phi_2, \phi_1; \phi_5, \phi_6} \Gamma_{\beta \beta}^{\alpha \alpha}(\mathbf{k}, \mathbf{p}^\prime; \mathbf{q})}{\epsilon_+ (\frac{\mathbf{q}}{2} - \mathbf{k}) + \epsilon_+ (\frac{\mathbf{q}}{2}+\mathbf{k})}
	+
	\frac{\mathcal{V}^{(5)}_{\phi_1, \phi_2; \phi_6, \phi_5} \Gamma_{\alpha \beta}^{\alpha \alpha}(\mathbf{k},\mathbf{p}^\prime; \mathbf{q})}{\epsilon_+ (\frac{\mathbf{q}}{2} - \mathbf{k}) + \epsilon_- (\frac{\mathbf{q}}{2}+\mathbf{k})}
	\right],
\end{align}
where $\Gamma_{\mu \nu}^{\rho \tau} (\mathbf{p}, \mathbf{p}^\prime; \mathbf{q})$ is the $t$ matrix for scattering of particles in the branches $\mu,\nu$ with momenta $\mathbf{q}/2 \pm \mathbf{p}$ to branches $\rho,\tau$ with final momenta $\mathbf{q}/2 \pm \mathbf{p^\prime}$ .
The key to solving this set of equations is to construct the quantities
\begin{align}
	&X(\mathbf{p}, \mathbf{p}^\prime; \mathbf{q})
	\equiv
	\frac{1}{4}
	\left(
	\Gamma_{\alpha \alpha}^{\alpha \alpha} (\mathbf{p}, \mathbf{p}^\prime ; \mathbf{q})
	+
	\Gamma_{\beta \beta}^{\alpha \alpha} (\mathbf{p}, \mathbf{p}^\prime ; \mathbf{q})
	\right.
	\notag \\
	&\left.
	+
	\Gamma_{\alpha \beta}^{\alpha \alpha} (\mathbf{p}, \mathbf{p}^\prime ; \mathbf{q})/\sqrt{2}
	+
	\Gamma_{\alpha \beta}^{\alpha \alpha} (-\mathbf{p}, \mathbf{p}^\prime ; \mathbf{q})/\sqrt{2}
	\right),
	\notag 
	\\
	&Y(\mathbf{p}, \mathbf{p}^\prime; \mathbf{q})e^{-i(\phi_3 + \phi_4)}
	\equiv
	\frac{1}{4}
	\left(
	\Gamma_{\alpha \alpha}^{\alpha \alpha} (\mathbf{p}, \mathbf{p}^\prime ; \mathbf{q})
	+
	\Gamma_{\beta \beta}^{\alpha \alpha} (\mathbf{p}, \mathbf{p}^\prime ; \mathbf{q})
	\right.
	\notag \\
	&\left.
	-
	\Gamma_{\alpha \beta}^{\alpha \alpha} (\mathbf{p}, \mathbf{p}^\prime ; \mathbf{q})/\sqrt{2}
	-
	\Gamma_{\alpha \beta}^{\alpha \alpha} (-\mathbf{p}, \mathbf{p}^\prime ; \mathbf{q})/\sqrt{2}
	\right)
	e^{-i(\phi_1 + \phi_2)},
	\notag 
\end{align}
and
\begin{align}
	&Z(\mathbf{p}, \mathbf{p}^\prime; \mathbf{q})
	\equiv
	\frac{1}{2}
	\frac{
	\Gamma_{\alpha \alpha}^{\alpha \alpha}(\mathbf{p}, \mathbf{p}^\prime; \mathbf{q})
	-
	\Gamma_{\alpha \alpha}^{\beta \beta}(\mathbf{p}, \mathbf{p}^\prime; \mathbf{q})
	}
	{(e^{i\phi_1} + e^{i\phi_2})(e^{-i\phi_3} + e^{-i\phi_4})}
	\notag \\
	&=
	\frac{1}{2}
	\frac{\Gamma_{\alpha \beta}^{\alpha \alpha} (\mathbf{p}, \mathbf{p}^\prime; \mathbf{q})/\sqrt{2}
	-
	\Gamma_{\alpha \beta}^{\alpha \alpha} (-\mathbf{p}, \mathbf{p}^\prime; \mathbf{q})/\sqrt{2}
	}{(e^{i\phi_1} - e^{i\phi_2})(e^{-i\phi_3} + e^{-i\phi_4})}.
\end{align}
Rewriting the Bethe-Salpeter equations in terms of $X$, $Y$, and $Z$, 
we see that $X$, $Y$, and $Z$ do not depend on their first arguments.
Namely, we can write $X(\mathbf{p}, \mathbf{p}^\prime ; \mathbf{q}) = X(\mathbf{p}^\prime; \mathbf{q})$, etc.
Then $X$, $Y$, and $Z$ inside the integrals can be moved outside, and we can solve for $X$, $Y$, and $Z$ algebraically,
and reconstruct $\Gamma_{\alpha \alpha}^{\alpha \alpha}$ from $X$, $Y$, and $Z$ via
\begin{align}
	&\Gamma_{\alpha \alpha}^{\alpha \alpha}(\mathbf{p}, \mathbf{p}^\prime; \mathbf{q})
	=
	X(\mathbf{p}^\prime; \mathbf{q}) + Y(\mathbf{p}^\prime; \mathbf{q}) e^{i(\phi_1 +\phi_2 - \phi_3 - \phi_4)}
	\notag \\
	&+Z(\mathbf{p}^\prime; \mathbf{q})(e^{i\phi_1} + e^{i\phi_2})(e^{-i\phi_3} - e^{-i\phi_4}).
\end{align}
Introducing the free field scattering lengths by $m/(4\pi a_{ij}) = 1/g_{ij} + m\Lambda / (2\pi^2)$, where $\Lambda$ is the high momentum cutoff, we obtain Eq.~(\ref{exactt}).

\end{document}